\documentclass[fleqn,usenatbib]{mnras}

\usepackage{newtxtext,newtxmath}

\usepackage[T1]{fontenc}
\usepackage{ae,aecompl}

\usepackage{graphicx}	
\usepackage{amsmath}	
\usepackage{amssymb}	

\title[32 years of optical spectroscopy for PU Vul]{An illumination effect and an eccentric orbit for the symbiotic binary PU Vul revealed by 32 years of optical spectroscopy}

\author[V. A. C\'uneo et al.]{
Virginia A. C\'uneo,$^{1,4}$\thanks{E-mail: vcuneo@iar-conicet.gov.ar}
Scott J. Kenyon,$^{2}$
Mercedes N. G\'omez,$^{3,4}$
Drahomir Chochol,$^{5}$
\newauthor 
Sergey Y. Shugarov$^{5,6}$
and Eugeni A. Kolotilov$^{6}$
\\
$^{1}$Instituto Argentino de Radioastronom\'ia (CCT La Plata, CONICET), C.C.5, (1984) Villa Elisa, Buenos Aires, Argentina\\
$^{2}$Smithsonian Astrophysical Observatory, 60 Garden Street, Cambridge, MA 02138, USA\\
$^{3}$Universidad Nacional de C\'ordoba, Observatorio Astron\'omico, Laprida 854, X5000BGR, C\'ordoba, Argentina\\
$^{4}$Consejo Nacional de Investigaciones Cient\'ificas y T\'ecnicas (CONICET), Argentina\\
$^{5}$Astronomical Institute, Slovak Academy of Science, 059 60 Tatransk\'a Lomnica, Slovakia\\
$^{6}$Sternberg Astronomical Institute, University avenue 13, Moscow 119992, Russia
}

\date{Accepted XXX. Received YYY; in original form ZZZ}

\pubyear{2018}

\begin{document}
\label{firstpage}
\pagerange{\pageref{firstpage}--\pageref{lastpage}}
\maketitle

\begin{abstract}
We analyze $\sim$32 years of optical spectra and photometry for the symbiotic binary PU Vul. Light curves for the He I $\lambda$4471, He II $\lambda$4686 and H$\beta$ $\lambda$4861 emission lines reveal an illumination effect, where the hot white dwarf ionizes the outflowing wind of the red giant, and evidence for an eccentric orbit with e $\geq$ 0.16. Along with the gradual appearance of high ionization emission from [Fe VII] and O~VI, the relative fluxes of these lines suggest an increase in the effective temperature of the hot component, from roughly $10^5$~K on JD 2448000 (1990) to roughly $2 \times 10^5$~K on JD 2455000 (2009). During this period, the luminosity of the hot component dropped by a factor of 4$-$6 to a current value of roughly 1000 L$_\odot$.
\end{abstract}

\begin{keywords}
binaries: spectroscopic -- binaries: symbiotic -- stars: individual:  PU Vul 
\end{keywords}


\section{Introduction}
Symbiotic stars are interacting binary systems where a main sequence, subdwarf, or white dwarf (WD) star accretes material from the wind of an evolved late-type star \citep[e.g.][]{1984ASSL..112.....A,1986syst.book.....K}. Based on the amplitude ($\Delta V$) and duration ($\Delta t$) of their outbursts at optical wavelengths, these systems are divided into a group of classical symbiotics ($\Delta V \lesssim$ 2--3~mag, $\Delta t \lesssim$ 1--2~years) and symbiotic novae ($\Delta V \approx$~3--6~mag, $\Delta t \gtrsim$ 10--100~years). Among the symbiotic novae, several (AG~Peg, RT Ser, RR Tel, and PU Vul) transform into A--F supergiants at optical maximum, gradually evolve towards hotter temperatures, and then develop strong emission lines similar to a planetary nebula \citep[e.g.,][]
{1980MNRAS.192..521A,1982ApJ...259..244I,1983ApJ...273..280K,1992MNRAS.256..177M,1996ApJS..105..145I}. In others (V1016 Cyg, V1329 Cyg, and HM Sge), the lifetime as an A--F supergiant is negligible; the strong nebular emission line spectrum appears either simultaneously with or almost immediately after optical maximum \citep{1994A&A...282..586M}.

PU Vul was discovered by Y. Kuwano \citep{1979IAUC.3344....1K} and M. Honda \citep{1979IAUC.3348....2A}. Before its slow rise in 1978-1979 (JD $\sim$2443660$-$2443969), the system varied between B~$=$~14.5 and B~$=$~16.6 \citep{1979AJ.....84.1357L}. At optical maximum, PU Vul reached V $\approx$ 9. In 1980 \mbox{(JD $\sim$2444532)}, the system fell almost 5 magnitudes in V, maintained this level for more than a year, and then recovered to \mbox{V $\approx$ 9}. After 7--8 years, the system began a slow optical decline interrupted by a second shallower minimum in 1994 \mbox{(JD $\sim$2449447)} and a third very shallow minimum observed only in the U-band in 2007 \citep[JD $\sim$2454334,][]{2012BaltA..21..150S}. As the luminosity declined, the radius of the WD shrunk from $\sim$0.35 R$_{\odot}$ to $\sim$0.04 R$_{\odot}$. Shortly after the second minimum, the system became faint enough to reveal pulsations in the red giant (RG) primary with a period of $\sim$218 days  \citep{1998IBVS.4571....1C,2012ApJ...750....5K,2012BaltA..21..150S}.

Throughout these epochs, optical and ultraviolet (UV) spectroscopic data provided windows into the evolution of the eruptive star. Shortly after outburst, the optical spectrum resembled an F-type supergiant \citep{1982PASJ...34..269Y,1991PASJ...43..225K}. Although this component disappeared during the 1980 minimum, it returned during the recovery to the second optical maximum. From 1982--1990, the spectrum evolved from an F-type star into an A-type star with occasional H~I Balmer emission lines, then to a middle B-type star with strong Balmer and weak Fe~II emission lines, and finally to a very blue star with broad He~II lines characteristic of Wolf-Rayet (WR) stars \citep{1989A&A...223..119B,1986AJ.....91..563K,1989A&A...215...57I,1991PASJ...43..523K,1991A&A...250..361G,1991MNRAS.252P..31T,1992A&A...259..525V,1993AJ....106.2118S,1994A&AS..104...99K}. During this transformation, red optical spectra showed prominent TiO absorption bands characteristic of M-type giant stars. Together with spectra acquired during the 1994 minimum, these data demonstrated that the system is a binary composed of a pulsating M6 giant and an eruptive WD. During the deep minima, the RG occults the WD and any surrounding ionized material. Thus, the three minima provide a rough estimate of the orbital period, 13.4~years \citep[][and references therein]{1984NASCP2349..305F,1986AJ.....91..563K,1996A&A...307..470N,1996JAVSO..24...81G,2011ARep...55..896T,2012ApJ...750....5K,2012BaltA..21..150S}. 

In current theory, symbiotic nova eruptions result from thermonuclear runaways in accreted material on the surface of the WD \citep[e.g.][]{1972ApJ...176..169S,2002ASPC..261..595K,2005ApJ...623..398Y}. During the decline from optical maxima, the WD should evolve in effective temperature from 5,000--6,000~K to 150,000--200,000~K at roughly constant luminosity and then evolve to lower luminosity at roughly constant radius. For PU Vul, detailed analyses of UV data appear to support this prediction \citep{2012ApJ...750....5K}: to within a factor of two, the hot WD maintained a constant luminosity from 1982--2011 as its effective temperature increased from 6,500~K to 150,000--165,000~K. However, optical spectra suggest a dramatic decline in luminosity, from $\sim$$10^{4}$~L$_{\odot}$ to $\sim$$10^{3}$ L$_\odot$ over 1992--2008 \citep{2009ARep...53.1020T,2011ARep...55..896T,tat2018}. Contrary to theoretical predictions, these data also imply an {\it increase} in effective temperature as the bolometric luminosity declined.

To try to resolve this difference, we analyze optical spectrophotometry of PU Vul acquired from 1984--2016 (JD~2445804--2457724). Aside from adding to our understanding of the evolution from an F-type supergiant into a WR-type star \citep[see also][]{1986AJ.....91..563K}, these data probe the evolution of emission lines during epochs when the luminosity remained roughly constant \citep{2012ApJ...750....5K} or declined by a factor of 10 \citep{2011ARep...55..896T}. The dense time-coverage of our spectra allows us to probe the evolution of the hot component on short and long time scales.

After summarizing the observations in \S\ref{obs}, we describe the general evolution of the spectrum, the evolution of fluxes for selected H and He emission lines, and a comparison with previous observations and theoretical predictions in \S\ref{evolution}. We then discuss the illumination effect, a new estimate of the orbital eccentricity, and the evolution of the temperature and luminosity of the hot component in \S\ref{analysis}. We conclude with a discussion of the main results in \S\ref{dis}.

\section{Observations}
\label{obs}
\subsection{Optical Spectroscopy}
Throughout 1984--92 (JD 2445804--2448935), S. K. acquired occasional low resolution optical spectra of PU Vul with the cooled dual-beam intensified Reticon scanner (IRS) or the GoldCam CCD dewar mounted on the white spectrograph of the KPNO No. 1 0.9-m telescope.  Observations of 5--10 standard stars place the spectra on the \citet{1975ApJ...197..593H} flux scale. Typical errors in the flux calibration are 3\% to 5\%. For the IRS, blue (3500-6200~\AA) and red (5800--8400~\AA) spectra acquired on consecutive nights yield an independent measure of the accuracy of the flux calibration.  For the region of overlap, the difference in the continuum level is small, $\lesssim$ 3\%. For these 21 spectra, the typical signal-to-noise in the continuum at 5500~\AA\ is 15--20 per pixel.

From 1994--2016 (JD 2449694--2457724), P. Berlind, M. Calkins, and several other observers acquired low resolution optical (3500-7300~\AA) spectra of PU Vul with FAST, a high throughput, slit spectrograph mounted at the Fred L. Whipple Observatory 1.5-m telescope on Mount Hopkins, Arizona \citep{1998PASP..110...79F}. They used a 300 g mm$^{-1}$ grating  blazed at 4750~\AA, a 3\arcsec~slit, and a thinned 512 $\times$ 2688 CCD. These spectra cover 3800--7500 \AA~at a resolution of 6 \AA. We wavelength-calibrate the spectra in NOAO IRAF\footnote{IRAF is distributed by the National Optical Astronomy Observatory, which is operated by the Association of Universities for Research in Astronomy, Inc. under contract to the National Science Foundation.}. After trimming the CCD frames at each end of the slit, we correct for the bias level, flat-field each frame, apply an illumination correction, and derive a full wavelength solution from calibration lamps acquired immediately after each exposure.  The wavelength solution for each frame has a probable error of $\pm$0.5 \AA~or better. To construct final 1-D spectra, we extract object and sky spectra using the optimal extraction algorithm APEXTRACT within IRAF. 

We acquired 458 FAST spectra with exposure times of 1--300 seconds. To estimate the signal-to-noise (S/N) in the continuum, we examined three different regions relatively free of emission lines ($4750$--$4840$ \AA, $5900$--$6070$ \AA, and \mbox{$6700$--$6800$ \AA}), measured fluctuations relative to the continuum level, and averaged the values for each spectrum.  Most spectra have moderate signal-to-noise, S/N $\gtrsim$ 15 per pixel, for an exposure time of 100 seconds. 

To flux-calibrate the PU Vul spectra on reasonably clear nights, we observed several standards from the IRS standard star manual \citep{1982book}. For each spectrum, we derive the extinction as a function of wavelength using the KPNO extinction curve and a polynomial interpolation routine from \citet{1992nrfa.book.....P}. After applying the extinction correction, we bin the counts in 50~\AA~ bins centered on the wavelengths in the NOAO standard star files included with IRAF releases (e.g., the "irscal" directory). Comparison with standard star magnitudes yields the correction factor for each bin. We use a polynomial interpolation routine \citep{1992nrfa.book.....P} to apply the binned correction factors to the original standard star spectrum. The resulting spectrum is then calibrated on the \citet{1975ApJ...197..593H} flux scale \citep[see also][]{1988ApJ...328..315M}. As a test of the flux-calibration, we integrate the corrected spectra over B and V filter response curves; our approach yields the correct B and V magnitudes for each standard to $\pm$0.01 mag. 

On any night, most standards are observed at airmasses within $\pm$0.1--0.2 of each PU Vul spectrum. To apply the flux-calibration to PU Vul spectra, we derive the median calibration for the set of standard stars. The inter-quartile range then provides an estimate of the error in the calibration. For nights with at least 4 standard star observations, the average and median calibrations agree to $\pm$0.02 mag. When nights are reasonably close to photometric, the inter-quartile range of the calibration is 0.02--0.03 mag. On non-photometric nights, the inter-quartile range is 0.1--1.5 mag. Table \ref{tablelog} lists the observing log; N is the number of individual exposures with a range of exposure times Etime. Sky conditions (Cond.) describe sky transparency as "good" when the error in the standard star observation is < 0.1 mag, "fair" when error $=$ 0.1-0.3 mag, and "poor" for error > 0.3 mag. The calibration is good for 81 nights, fair for 52 and poor for 22. We see that at least half of the observations need a better calibration, we decided then to use photometry from \citet{2012BaltA..21..150S} and new data acquired by the authors to minimize the systematic error, as explained in Section \ref{lineflu}.

\begin{table}
\caption{Log of observations. We include here only the information for the first ten nights in our sample. The full version of the table is available as online-only material.}
\label{tablelog}
\centering
\begin{tabular}{cccc}
\hline 
JD & N & Etime & Cond. \\
\hline
2445804 & 1 & 480 &	good \\
2445985 & 1	& 600 &	good \\
2446222 & 1	& 480 &	good \\
2446723 & 1	& 1200 & good \\
2447108 & 1	& 720 &	good \\
2447496 & 1	& 720 &	good \\
2448408 & 1	& 920 &	good \\
2448526 & 1	& 960 &	good \\
2448935 & 4	& 150 &	good \\
2449694	& 2 & 5--60 & good \\
\hline
\end{tabular}
\end{table}

\subsection{Optical Photometry}
To improve the calibration of the FAST spectra, we compare B and V magnitudes from our FAST calibration with an extensive set of B and V photometry. As described in \citet{2012BaltA..21..150S}, contemporaneous broadband UBVRI photoelectric and CCD data acquired at three different observatories were transformed to a common photometric system, UBVR$_J$I$_J$. The photometric data in \citet{2012BaltA..21..150S} cover the period 1979--2011. For FAST spectra acquired after 2011, we make a comparison with new photometry from the same observatories using the same acquisition and reduction procedures described in \citet[][and references therein]{2012BaltA..21..150S}. The new photometry is included at the Table ``new\_photometry.txt", available as online-only material. Overall, these data have a typical uncertainty of \mbox{0.01--0.02 mag}.

\section{Spectroscopic Evolution of PU Vul: 1984-2016}
\label{evolution}
\subsection{General Evolution of the Optical Spectrum}
\begin{figure*}
\centering
\includegraphics[trim=0 7mm 0 0,clip,width=\textwidth]{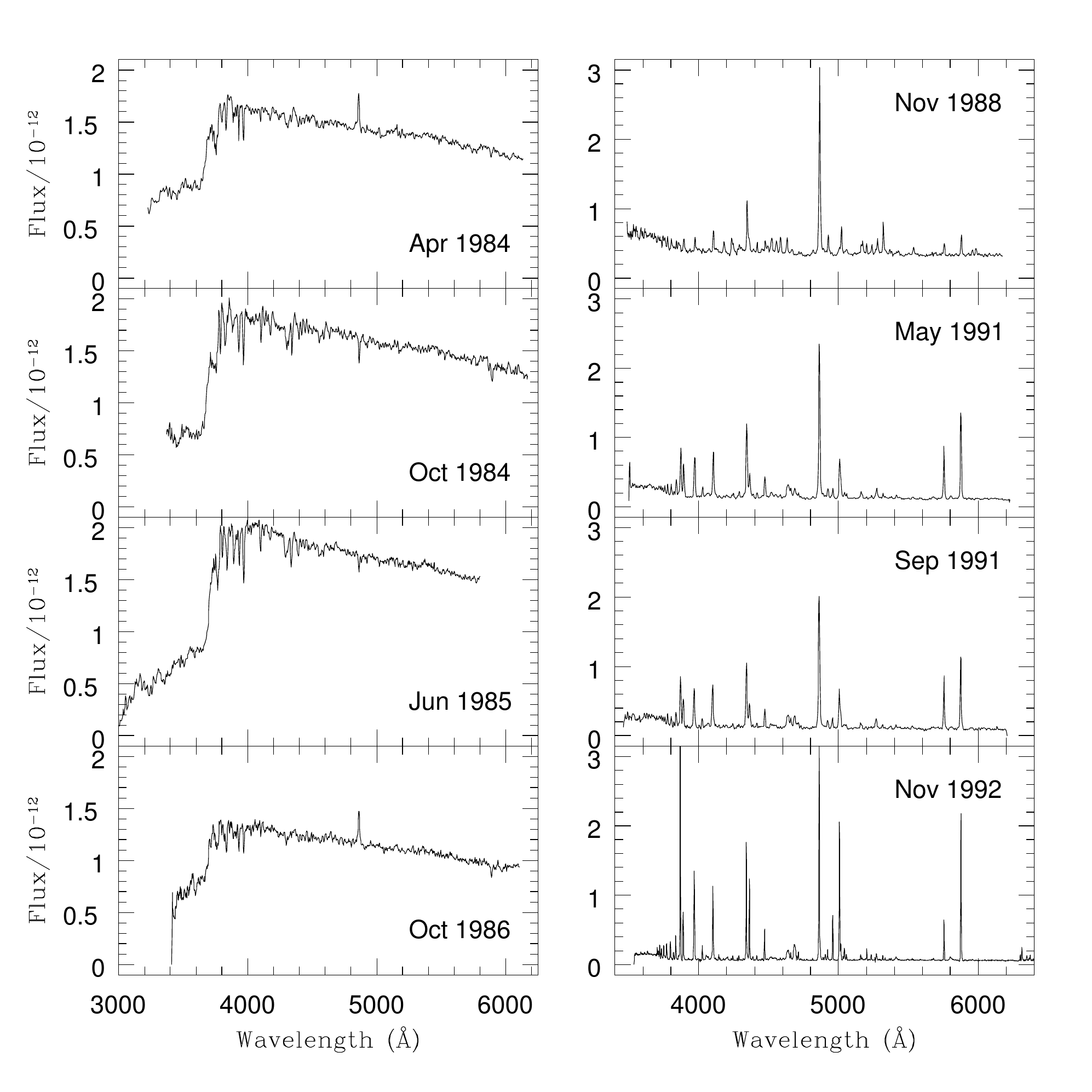}
\caption{Sample of the PU Vul blue spectra obtained between 1984 (JD 2445804) and 1992 (JD 2448935). Fluxes are in units of erg s$^{-1}$ cm$^{-2}$ \AA$^{-1}$.
\label{fig:spec_1}}
\end{figure*}

To develop a better understanding of the changing optical spectrum, we first describe the long-term behavior of absorption and emission features. During 1984--92 (JD 2445804 --2448935), the optical spectrum of PU Vul evolved dramatically on the IRS spectra (Figure~\ref{fig:spec_1}). Prior to 1987 (JD~$\sim$2446723), the system resembled an A--F supergiant with occasional H$\alpha$ and H$\beta$ emission lines \citep[see also][]{1986AJ.....91..563K,1991MNRAS.249..374I}. As the A--F absorption features vanished in 1987--88 (JD $\sim$2447496), the blue continuum strengthened; prominent He~I and Fe~II/[Fe~II] emission lines also appeared \citep[see also][]{1989A&A...215...57I,1991A&A...250..361G,1991PASJ...43..523K,1994A&AS..104...99K}. In 1991--1992 (JD~2448408--2448935), the He~I lines reached a maximum intensity, $\sim$2--3 times stronger than a much broader He~II $\lambda$4686 feature. As the He~I lines intensified, the standard [O~III] emission lines at $\lambda$4363, 4959, 5007 also strengthened, with $\lambda$4363 ($\lambda$5007) at roughly 75\% (50\%) of the intensity of H$\gamma$ (H$\beta$). In September 1991 (JD 2448526), the [N~II] $\lambda$5755 nebular line peaked at nearly 80\% of He~I $\lambda$5876.

\begin{figure*}
\centering
\includegraphics[trim=0 7mm 0 0,clip,width=\textwidth]{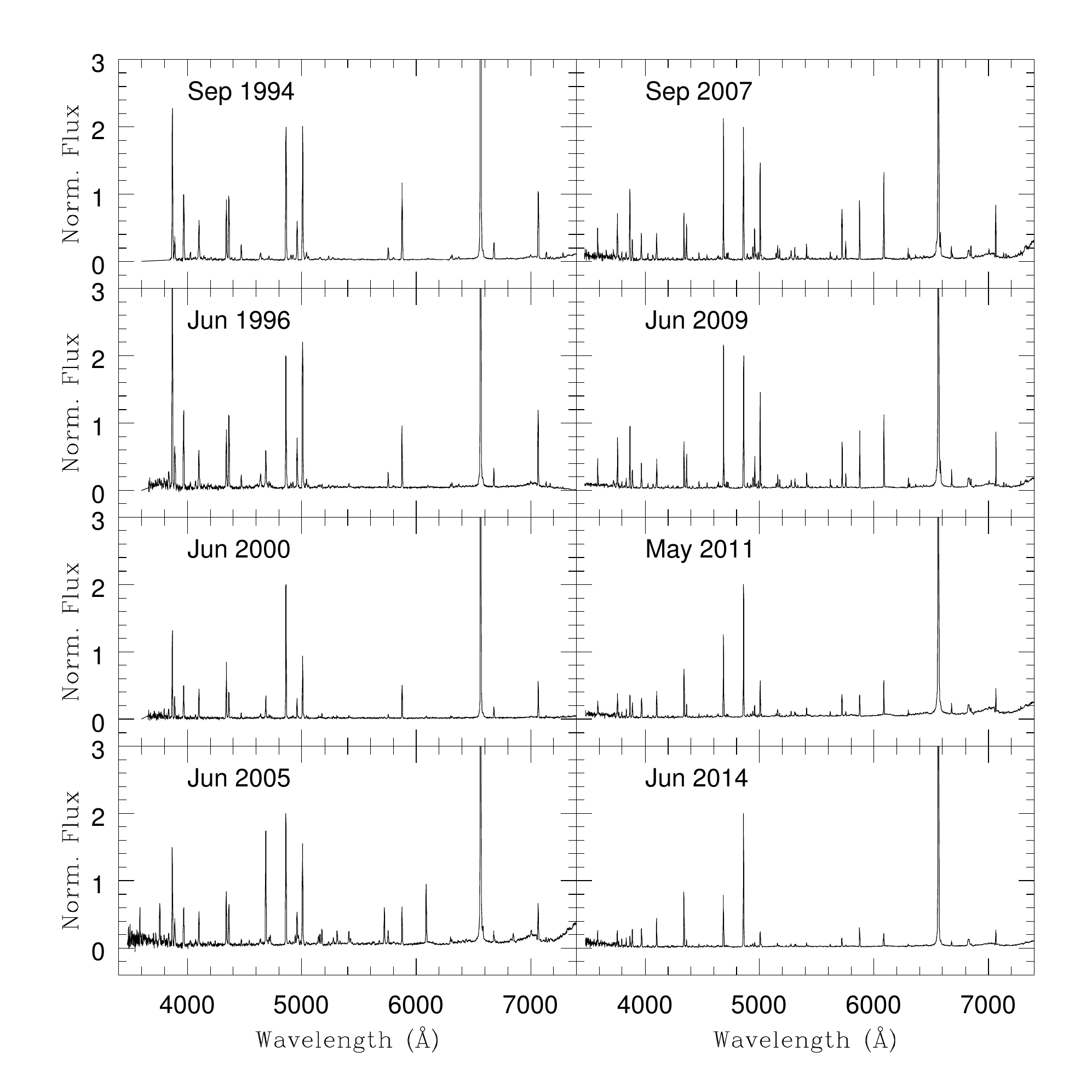}
\caption{Sample of the PU Vul spectra obtained between 1994 (JD~2449609) and 2014 (JD~2456833). Normalized fluxes are fluxes relative to H${\beta}$, normalized
to a value of 2.\label{fig:spec}}
\end{figure*}

To summarize the evolution in 1994--2016 (JD~2449694--2457724), we rely on the FAST spectra (Figure \ref{fig:spec}).  During 1994--1996 (JD~2449609--2450258), [O~III] emission surpassed H$\beta$ and H$\gamma$. By June 2000 (JD 2451701), however, H$\beta$ and H$\gamma$ were much stronger than neighboring [O~III] lines. As [O~III] emission weakened, He~II $\lambda$4686 strengthened; relative to H$\beta$, the He~II $\lambda$4686 flux grew by more than a factor of 10 by 2005 (JD~2453525). In addition to the growth in He~II $\lambda$4686 emission, the June 2005 (JD 2453525) spectrum was much richer than previous spectra, with strong emission from [Fe~VII] $\lambda3758$ and $\lambda$6087 and the Raman scattered O~VI band at $\lambda$6830. Although the emission line fluxes continue to evolve after 2005, the high excitation features remained prominent on all of the FAST spectra \citep[see also][]{tat2018}.

\begin{figure}
\includegraphics[trim=10mm 7mm 101mm 7mm,clip,width=\columnwidth]{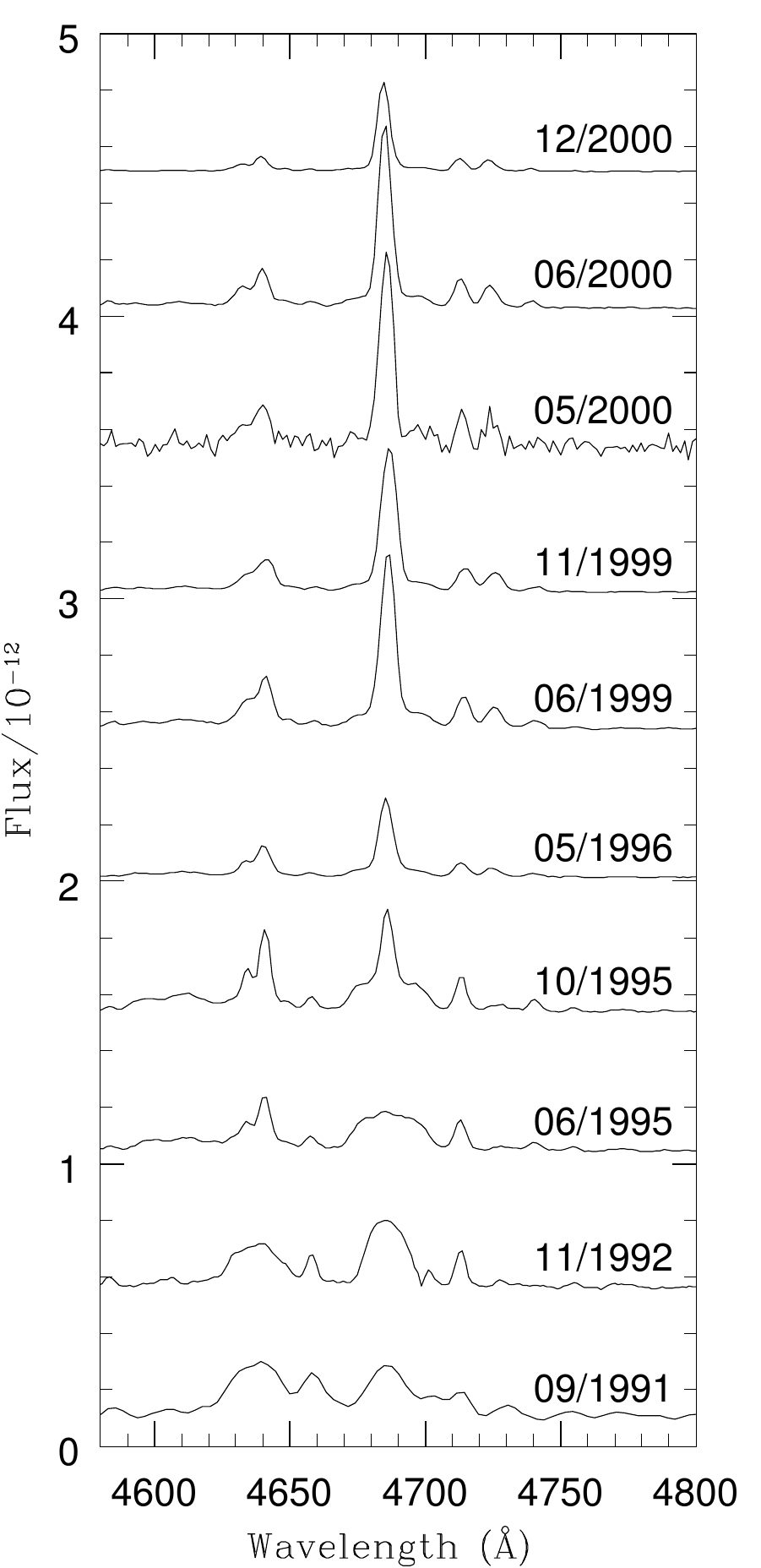}
\caption{Evolution of He~II $\lambda$4686 between 1991 (JD~2448526) and 2000 (JD~2451903). Each spectrum was displaced a factor multiple of 0.5 from the flux base level to a better display. Fluxes are in units of erg s$^{-1}$ cm$^{-2}$ \AA$^{-1}$.\label{fig:4686}}
\end{figure}

To focus on the development of WR features in 1991--2000 (JD~2448526--2451903), Figure \ref{fig:4686} shows the evolution of He~II $\lambda$4686 and the N~III/C~III blend near $\lambda$4640. Early on, these features were weak but broad, suggestive of outflow with a velocity of approximately 1000~km~s$^{-1}$ \citep[see also][]{1991MNRAS.252P..31T}. As these broad lines maintained a roughly constant equivalent width from 1991--1995 (JD~2448526--2450019), narrow emission features grew stronger. Although the broad features remained visible until 2001 (JD~2452175), they were much weaker relative to the narrow components.  We do not have spectra between 2001 (JD~2452175) and 2005 (JD~2453525), but by 2005 the broad features were below our detection limit and remained undetectable through 2016 (JD~2457724). 

Other WR-type emission lines -- e.g., faint N~IV $\lambda$4058 \citep{1991MNRAS.252P..31T} -- were also weakly present from May 1991 (JD 2448408) until November 1992 (JD 2448935) but were not detected during 1994--2016 (JD~2449694--2457724). Because our spectra have fairly low resolution, it is possible that these features were present but masked by numerous narrow features.

Aside from \citet{1991MNRAS.252P..31T}, \citet{1995IBVS.4251....1A} reported the evolution of WR features on spectra acquired during 1991--1995 \citep[see also][]{1995MNRAS.275..185K}. All of the broad emission lines disappeared during the 1994 minimum and re-appeared on post-eclipse spectra. Although we do not have spectra covering the minimum, He~II $\lambda$4542 and C~IV emission at $\lambda$5801 and $\lambda$7226 were visible after the eclipse. Broad C~IV $\lambda$7226 emission disappears during 1999--2000 (JD~2451335--2451903); $\lambda$5801 disappeared definitely by 2005 (JD~2453525). After the eclipse, \citet{1995IBVS.4251....1A} also note the development of a narrow emission feature superposed on the broader WR components.  They measured a displacement of 150~km~s$^{-1}$ between the centroids of the narrow and broad features. On our low resolution spectra, it is impossible to measure any shift between the two components.

\subsection{Line fluxes and the light curve}
\label{lineflu}
To derive equivalent widths and line fluxes, we used the \textit{splot} task within IRAF. After we choose by eye two points on either side of the lines for tracing the continuum, the routine fits a gaussian profile to the line intensity. We also used the \textit{sbands} task, specifying the central wavelength of the line, two well-spaced points to define the continuum, and an adequate band width to measure the line flux. Comparing the values obtained with both methods, the differences are within the measurement errors. The choice of the continuum is the main source of error. In the following discussion, we adopt fluxes measured with \textit{splot}, which provides more flexibility in defining the continuum. 

On most nights, we have a set of three FAST spectra with different exposure times. Short exposures avoid saturating the strongest lines but have poor S/N for the rest of the spectrum. Intermediate and long exposures yield optimal S/N for moderate and weak lines. For each line in this study, we adopted equivalent widths and fluxes from spectra with the longest exposure time which did not saturate the feature. Thus, we have one optimal measurement per feature, per night. As mentioned previously, roughly half of the spectra were acquired on nights with significant cirrus. To place all of the spectra on the same photometric scale, we normalize the fluxes to the optical broadband photometry from \citet{2012BaltA..21..150S}. We interpolated the BV photometry to the dates of our FAST observations and corrected the line fluxes for He~I $\lambda4471$, He~II $\lambda4686$, and H$\beta$ $\lambda4861$ using the difference in B magnitudes, $\Delta B = B(FAST) - B(phot)$. Tests using a correction derived from B and V photometry interpolated to the wavelengths of the three lines yielded nearly identical results. Thus, we employed the simpler correction based on interpolated B photometry to place all of the FAST line fluxes on the proper photometric scale. On some nights of very poor quality, the photometric calibration failed. We discarded these data. Table \ref{table} lists the photometric calibrated fluxes for the H$\gamma$ $\lambda4341$, He I $\lambda4471$, He II $\lambda4686$, H$\beta$ $\lambda4861$ and H$\alpha$ $\lambda6563$ lines. 

\begin{table}
\caption{Line fluxes in units of $10^{-12}$ erg s$^{-1}$ cm$^{-2}$. We include here only the line fluxes for the first ten spectra in our sample. The full version of the table is available as online-only material.}
\label{table}
\begin{tabular}{cccccc}
\hline 
JD & H$\gamma$ & He I $\lambda 4471$ & He II $\lambda 4686$ & H$\beta$ & H$\alpha$\\
\hline
2445804 &  1.98 & 0.00 & 0.00 &  4.61 &  45.37 \\
2445985 &  1.08 & 0.03 & 0.00 &  2.70 &  26.10 \\
2446222 &  0.58 & 0.00 & 1.04 &  1.52 &  16.58 \\
2446723 &  1.67 & 0.00 & 0.00 &  4.63 &  48.90 \\
2447108 &  7.34 & 0.05 & 0.00 & 19.83 & 202.51 \\
2447496 & 12.25 & 0.08 & 0.00 & 29.88 & 294.75 \\
2448408 &  8.27 & 0.14 & 1.44 & 23.63 & 227.60 \\
2448526 &  9.35 & 0.12 & 2.30 & 22.00 & 210.21 \\
2448935 &  8.46 & 0.14 & 4.51 & 22.27 & 205.45 \\
2449694 &  1.66 & 0.62 & 0.79 &  4.04 &  39.55 \\
\hline
\end{tabular}
\end{table}

The line fluxes we measured have several sources of error. By using the photometry to rescale fluxes, we have (i) a small error in the photometry, $\sim$0.01--0.02 mag, (ii) another modest, $\sim$0.01--0.02 mag, error from the S/N of the spectra ($\sim$10--20) integrated over the 10 pixel width of the line, (iii) a $\sim$0.05 mag error in the placement of continuum due to the modest S/N of the spectra, and (iv) the error from blends in the line or any extra flux in a broad component. For these lines, blends are negligible; however, an unidentified broad component could contribute $\sim$5$\%$ to 10$\%$ of the total flux. Altogether, the errors in the fluxes are $\sim$10$\%$ to 15$\%$ with the dominant sources of uncertainty as error in the continuum fit and error in the flux from a broad component.

\begin{figure*}
\centering
\includegraphics[trim=0 28mm 0 28mm,clip,width=0.9\textwidth]{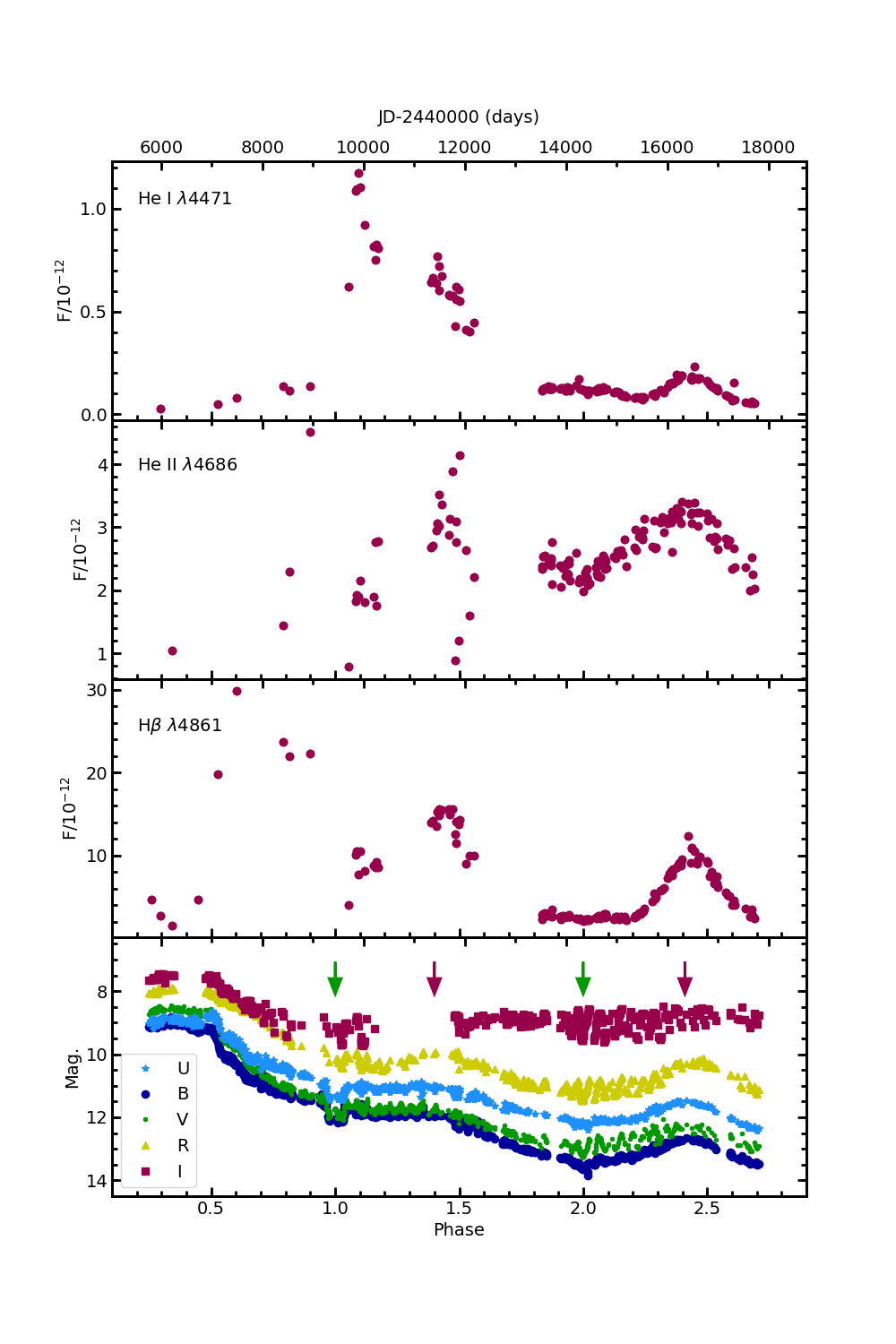}
\caption{Fluxes as function of phase and time for He I $\lambda4471$, He II $\lambda4686$ and H${\beta}$ $\lambda4861$ lines (top three panels) and the light curve (bottom panel), in the U, B, V, R and I bands, of PU Vul. Fluxes are in units of erg 
s$^{-1}$ cm$^{-2}$. The green arrows point the second and third eclipses at phases 1 and 2, respectively; the purple arrows denote the epochs of maximum light at phases $\sim$1.4 and $\sim$2.4.
\label{fig:flux_lc}}
\end{figure*}

The top three panels of Figure \ref{fig:flux_lc} show the photometry corrected fluxes for the He I $\lambda4471$, He II $\lambda4686$ and H$\beta$ $\lambda4861$ lines as a function of orbital phase and time for the period between April 1984 (JD 2445804) and December 2016 (JD 2457724). The orbital phase was derived from the ephemeris JD$_{Min}=2444537+4901\times E$, a linear fit to the three determined primary minima at JD 2444532, 2449447 and 2454334 \citep{2012ApJ...750....5K,2012BaltA..21..150S}.

The bottom panel of Figure \ref{fig:flux_lc} shows the light curve of PU Vul in the U, B, V, R and I bands, for the same time period. Within these light curves, the midpoints for eclipses 2 (1994) and 3 (2007) are denoted with green arrows. Aside from the eclipses and the abrupt decline starting around phase 0.5, there are several clear features in the light curve: (i) a sinusoidal oscillation in the brightness with maxima denoted by purple arrows (and slightly to the right for the R-band) and (ii) variations generated by the pulsation of the RG, which are more prominent after the eclipse at phase 2.0 \citep[see, for example,][]{2012BaltA..21..150S}. 

Despite the lack of spectra before 1994 (JD~2449694), the rise in H$\beta$ and He I fluxes is clearly associated with the drop in optical brightness starting at phase 0.5. The H$\beta$ line seems to reach a peak sometime during phase 0.6--0.8, while the He I $\lambda4471$ line shows a maximum close to phase 1.1, after the eclipse. From our limited phase coverage during this epoch, there is little evidence for an eclipse in any of the three lines. After the eclipse, He I emission declines, reaching a clear minimum sometime during phase 1.75--2. However, there is a small increase in emission that coincides with the optimal maximum near phase 2.5.

The H$\beta$ line (i) recovers from its decline to reach a second maximum which coincides with the maximum in the broadband light curve around phase 1.4, (ii) declines to a minimum around phase 2.0, and (iii) recovers again to a third maximum when the broadband light is bright just before phase 2.5. Comparing the three maxima in H$\beta$, there appears to be a decline in flux from the first maximum to the second and third maxima. However, this decline is small compared to the overall fluctuation in H$\beta$ emission. 

For the He II $\lambda$4686 line, a similar behavior is observed. Due to the lack of data before the eclipse, we can not say much about the behavior of the line. But, as for H$\beta$, there seems to be a maximum around phase 1.4. Then the flux declines reaching a minimum at phase 2, which coincides with the third eclipse. And finally, the flux rises again reaching a maximum before phase 2.5. Long-term fluctuations in the R-band are clearly phased with variations in the three lines.

\begin{figure}
\includegraphics[trim=15mm 121mm 20mm 15mm,clip,width=\columnwidth]{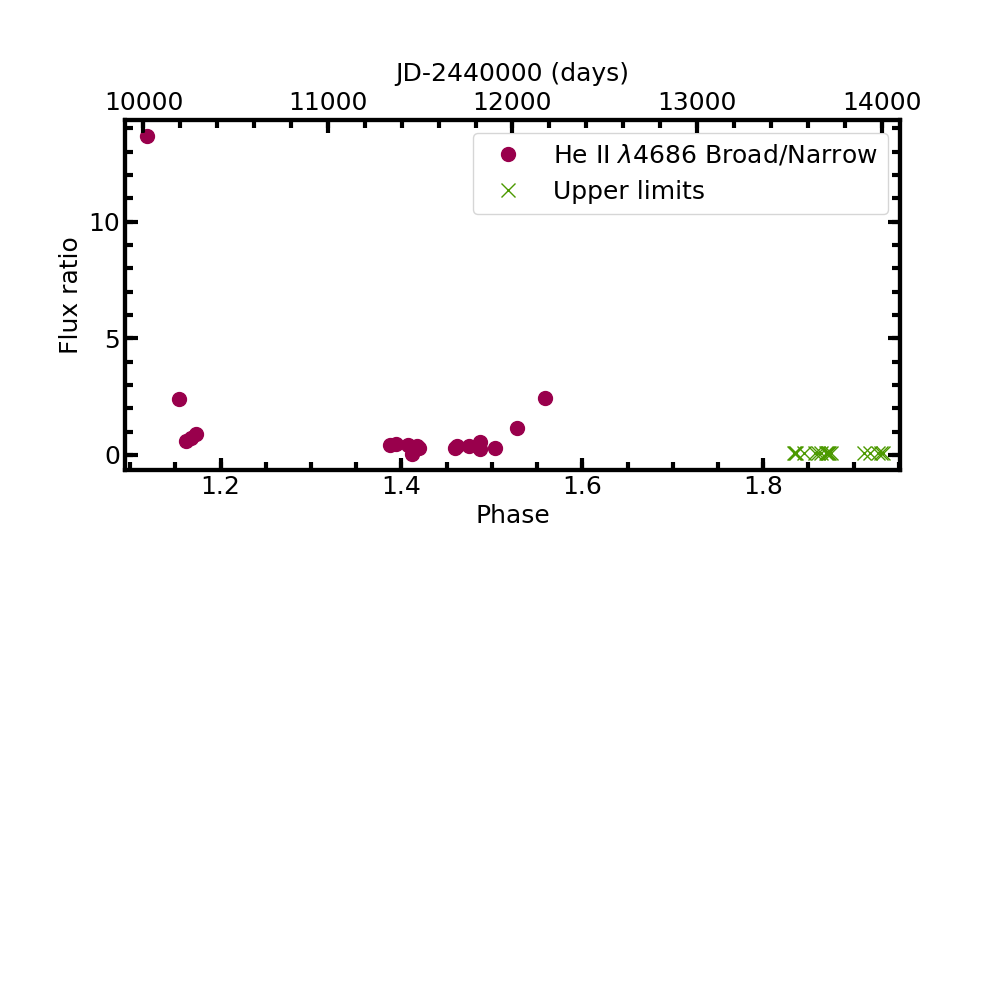}
\caption{Broad to narrow component ratio for the He II $\lambda4686$ line flux between JD $\sim$2450018 (October 1995) and JD $\sim$2454006 (September 2006). \label{fig:ratio}}
\end{figure}

Between $\sim$1995 (JD $\sim$2450018) and $\sim$2001 (JD~$\sim$2452175), He II $\lambda$4686 is composed of a broad and a narrow component. Prior to 1995, the line is either absent or consists only of the broad component. After 2001, only the narrow component is visible. For the period with two components, we fit two gaussians to derive the relative contribution of each component to the total flux (see Table \ref{table_4686}). Figure \ref{fig:ratio} shows the flux ratio of the broad to the narrow component derived from the two-gaussian fits. Aside from two spectra near JD $\sim$2452000, the broad component gradually weakened relative to the narrow component. Once our spectroscopic observations resumed after JD $\sim$2453525, the ratio of the broad to narrow component is always smaller than $\sim$0.1--0.2 and sometimes close to zero. Due to the gap in our data during JD $\sim$2452175--2453525, we cannot know when the broad component weakens after JD $\sim$2452175.

\begin{table}
\centering
\caption{He II $\lambda4686$ broad and narrow fluxes, normalized to the B-band photometry.}
\begin{tabular}{ccc}
\hline 
JD & F$_{Broad}$ & F$_{Narrow}$ \\
 & [$\times 10^{-12}$ erg s$^{-1}$ cm$^{-2}$] & [$\times 10^{-12}$ erg s$^{-1}$ cm$^{-2}$]\\
\hline
2450019 & 1.696 & 0.124 \\
2450192 & 1.337 & 0.556 \\ 
2450231 & 1.048 & 1.715 \\
2450258 & 1.406 & 1.925 \\
2450284 & 1.300 & 1.474 \\
2451335 & 0.773 & 1.901 \\
2451369 & 0.978 & 2.028 \\
2451431 & 0.876 & 2.074 \\
2451454 & 0.097 & 2.972 \\
2451484 & 0.964 & 2.558 \\
2451495 & 0.819 & 2.694 \\
2451690 & 0.669 & 2.343 \\
2451701 & 0.827 & 2.309 \\
2451762 & 0.989 & 2.585 \\
2451822 & 1.119 & 1.967 \\
2451824 & 0.573 & 2.190 \\
2451903 & 0.936 & 3.206 \\
2452026 & 1.331 & 1.136 \\
2452175 & 1.741 & 0.709 \\
\hline
\end{tabular} 
\label{table_4686}
\end{table}

Overall, the broad component of He~II $\lambda4686$ has a width consistent with the WR features observed in other symbiotic stars (Figure \ref{fig:4686}). Once it became visible, the narrow component of the He II $\lambda$4686 line has a width similar to H$\beta$. On all our spectra, there is no evidence for broad features on He I $\lambda4471$ or H$\beta$. Thus the broad components of these two lines have fluxes less than 10\% of our measured fluxes. 

\subsection{Comparison with other works}
From the tight variation of the H$\beta$ and He~I $\lambda$4471 lines at orbital phases 1.8--2.7, the relative calibration of our line fluxes appears excellent. The He~II $\lambda$4686 line shows more scatter in this period, but the scatter is still small relative to the amplitude of the variations. To judge the accuracy of the absolute calibration, we consider comparisons with other investigators. Although data from \citet{2011ARep...55..896T} cover a much shorter time period (2001$-$2008, JD~2452167--2454763), results for nights reasonably close in time yield flux ratios $F_T / F_0$ = 4.22$\pm$2.74 (He~I $\lambda$4471), 1.61$\pm$0.12 (He~II $\lambda$4686) and 1.71$\pm$0.18 (H$\beta$), where $F_T$ is the flux listed in Table~3 of \citet{2011ARep...55..896T} and $F_0$ is our measurement. Despite the quoted 10\% to 15\% uncertainty in both sets of fluxes, our measurements are roughly 60\% to 70\% of theirs. Still, the flux ratios are consistent within the 3$\sigma$ uncertainties.  Compared to data for He~II $\lambda$4686 in \citet{1995IBVS.4251....1A}, our fluxes are roughly 50\% larger before the eclipse of 1994 and a factor of 2--3 smaller after the eclipse. In both cases, it is possible that significant week-to-week or month-to-month variations complicate comparisons of different data sets.

As another test, we consider the flux ratios of several H and He lines. For the H~I Balmer lines, case B recombination predicts ratios of 0.45--0.5 for H$\gamma$/H$\beta$ and 2.7--3 for H$\alpha$/H$\beta$ \citep{1989agna.book.....O}. With $E(B-V)$ = 0.3, our results are 0.41$\pm$0.04 for H$\gamma$/H$\beta$ and 7.3$\pm$0.5 for H$\alpha$/H$\beta$. As is typical for symbiotic stars, the data suggest a modest optical depth -- $\tau \approx$ 5--10 in the H$\alpha$ line. Our results for the ratio of He~II $\lambda5411$ ($\lambda$4542) to He~II $\lambda$4686, 0.085$\pm$0.003 (0.041 $\pm$ 0.013) are identical to the predicted 0.08 (0.036). Finally, published fluxes for the He~II $\lambda$1640 lines \citep{2009ARep...53.1020T,2011ApJ...727...72K} -- dereddened with $E(B-V)$ = 0.3 \citep{1979ARA&A..17...73S,2012ApJ...750....5K} -- yield $F(\lambda1640) / F(\lambda4686)$ = 4.44$\pm$1.23, which is reasonably close to the predicted case B value of 6--7 \citep{1984ASSL..112.....A}. \citet{1987ApJS...63..965H} noted that the optical depth in the He~II lines of WR stars often yields ratios far from the case B value. The large variation in the ratio of He~II $\lambda$1640 to $\lambda$4686 in PU Vul is probably also due to large optical depth.

Finally, we compare the predicted nebular continuum emission (from He~II $\lambda$4686 and H$\beta$) with the observed continuum levels from the FAST spectra and the UBVRI photometry. From case B recombination theory, the predicted continuum level is a simple function of the line fluxes, the electron density $n_e$, and the electron temperature $T_e$ \citep{1984ASSL..112.....A,1989agna.book.....O}. For $n_e \gtrsim 10^6$~cm$^{-3}$ and $T_e \approx$ 10,000--20,000~K, we derive the predicted continuum flux at 4400~\AA\ for each set of line fluxes. Over the complete set of FAST spectra, the predicted continuum fluxes are roughly 25\% to close to 100\% of the observed continuum level. Following the same procedure for 3500--4000~\AA, the predicted level is again a significant fraction of the observed U-band flux. The observed line fluxes are therefore consistent with the observed continuum level at short wavelengths. At longer wavelengths (e.g., V and R bands), the predicted nebular continuum is much smaller. Considering the much larger contribution of the RG to the observed continuum (see below), the predicted nebular continuum is also consistent with the observations at these wavelengths.

\section{Analysis}
\label{analysis}
\subsection{Illumination effect and orbital eccentricity}
For phases 1.8--2.7, the time variation of the broadband light curve and the optical line fluxes shows clear evidence for an illumination effect, where high energy photons from the hot WD illuminate the outer atmosphere and the outflowing wind of the RG by photoionizing H, He and other elements, which then recombine to produce line and continuum emission \citep{1984ApJ...284..202S,1987A&A...182...51N,1996ApJ...471..930P,1998ApJ...501..339P}. Near phase 2, the dips in the U light curve and the He II $\lambda$4686 flux suggest an eclipse, which is observed more clearly at phase 1. During these epochs, the RG occults the hot WD and the surrounding nebula. Also, as the densest part of the ionized wind lies between the hot component and the RG, when the RG is in front of the WD it occults a portion of this region, so the observed flux is low. Because the WD is small and the ionized nebula is optically thin, the hot component cannot occult the RG. Roughly 40\% of an orbital cycle after the eclipse of the WD, when the WD lies in front of the RG, there is a clear rise in the line fluxes and the broadband optical light from recombination. The R-band light curve has two clear peaks near phases 1.4 and 2.4. Although our spectroscopic coverage is limited, there seems to be a maximum in H$\beta$ and perhaps the other lines at phase 1.4. The peaks at phase 2.4 are obvious. 

This behavior in symbiotic stars is fairly typical \citep[e.g.,][]{1968AZh....45..139B,1982PASP...94..165K,1984PASP...96..321K,1990A&A...227..121F,2008JAVSO..36....9S,2016AJ....152....1K}. As discussed in \citet{1996ApJ...471..930P,1998ApJ...501..339P} and \citet{2008JAVSO..36....9S}, illumination of the RG photosphere is insufficient to generate an observable reflection effect. The RG simply does not cover enough area to intercept a significant fraction of the ionizing photons emitted by the hot WD. If the RG has an outflowing wind, however, this material is optically thick to high energy photons. The wind has a large surface area and can absorb 25\% to more than 50\% of the radiation emitted by the WD. Recombinations in the ionized wind produce Balmer continuum emission and strong emission lines. When the hot WD lies in front of the RG and its wind, Earth-bound observers see strong peaks in the broadband light curves and the emission line fluxes. Nearly half an orbital phase later, the illuminated hemisphere of the RG and its wind now lie behind the RG, which produces a strong minimum in lines and the continuum. 

In a binary with a circular orbit, the maximum from illumination occurs half an orbital phase after any eclipse \citep[e.g.,][]{2008JAVSO..36....9S,2016AJ....152....1K}. When the orbit is eccentric, the peak from illumination is offset from phase 0.5. \citet{2016AJ....151..139M} derived a relation among the eccentricity ($e$), the longitude of the periastron ($\omega$) and the eclipse timings for eclipsing binaries. In terms of the orbital phase, for $i\simeq 90$\textdegree, they obtained

\begin{equation}
e~ cos(\omega)=\frac{\pi}{2}(\phi_{S}-\phi_{P}-0.5), \label{ecce}
\end{equation}

\noindent where $\phi_{P}$ and $\phi_{S}$ are the phases for primary and secondary eclipses, respectively. If $\omega=$ 90 or 270\textdegree, when the observer views the system along the major axis of the orbit, equation \ref{ecce} suggests the phase difference between successive eclipses is 0.5 for any eccentricity. But for different values of $\omega$ the situation changes. In Figure \ref{fig:flux_lc} we observe clearly that the primary eclipses for PU Vul, when the RG occults the WD, occur for phases 1 and 2. The secondary eclipses take place when we observe a maximum in the light curve, at phases $\sim$1.4 and $\sim$2.4. As we explained previously, we observe a maximum instead of a minimum because the light comes mainly from the WD, that has a negligible size compared to the RG, and because of the reflection effect taking place where the WD illuminates the RG. Then, considering these phases in equation \ref{ecce}, the offset of the illumination peak suggests the orbit in PU Vul has an eccentricity of $\approx$ 0.16, when $\omega=$ 0 or 180\textdegree. This value represents a lower limit for the eccentricity, which depends on $\omega$.

\subsection{Temperatures of the hot component}
Methods for deriving the temperature of the hot component from optical emission lines rely on assumptions about the geometry of the ionized nebula. In a powerful method originally developed by \citet{ambar1932} and modified by \citet{1981psbs.conf..517I} to include the He~I $\lambda$4471 transition, the fluxes for He~I $\lambda$4471, He~II $\lambda$4686, and H$\beta$ serve as proxies for the number of H-ionizing and He-ionizing photons. The method assumes that the nebula is optically thick to ionizing photons with $\lambda <$ 912 \AA; photons with $\lambda <$ 912, 504 and \mbox{228 \AA} ionize H$^0$, He$^0$ and He$^+$, respectively \citep[see Fig. 2.2 at][]{1989agna.book.....O,1964MNRAS.127..217H}. Adopting a blackbody radiation source and electron temperatures of 10,000 K, the temperature of the hot component is

\begin{equation}
T_{h}\times 10^{-4} = 19.38\left(\frac{2.22F_{4686}}{4.16F_{H_{\beta}}+9.94F_{4471}}\right)^{1/2}+5.13. \label{iij}
\end{equation}

This method is valid for $70,000<T_{h}<200,000$ K. Aside from a central blackbody, this relation assumes that the lines form in regions with a similar geometry. If the lines are optically thin, the fluxes are independent of geometry. However, we measured that H$\alpha$ is optically thick with a flux relative to H$\beta$ that varies with time. The variation in this flux ratio implies a variation of the optical depth through the nebula, which may result in variations in the line fluxes with orbital phase. For PU Vul, several observations imply the assumption of line formation in regions with a similar geometry is incorrect: (i) when He~II $\lambda$4686 has a broad component, He~I and H~I lines do not and (ii) the variation of He~II $\lambda$4686 with orbital phase differs from the variations of the He~I and H~I lines. Still, this method gives us an idea of the variation of $T_h$ with time for comparison with results from UV data \citep{2011ApJ...727...72K,2012ApJ...750....5K}.

\begin{figure*}
\centering
\includegraphics[trim=0mm 20mm 0mm 6mm,width=\textwidth]{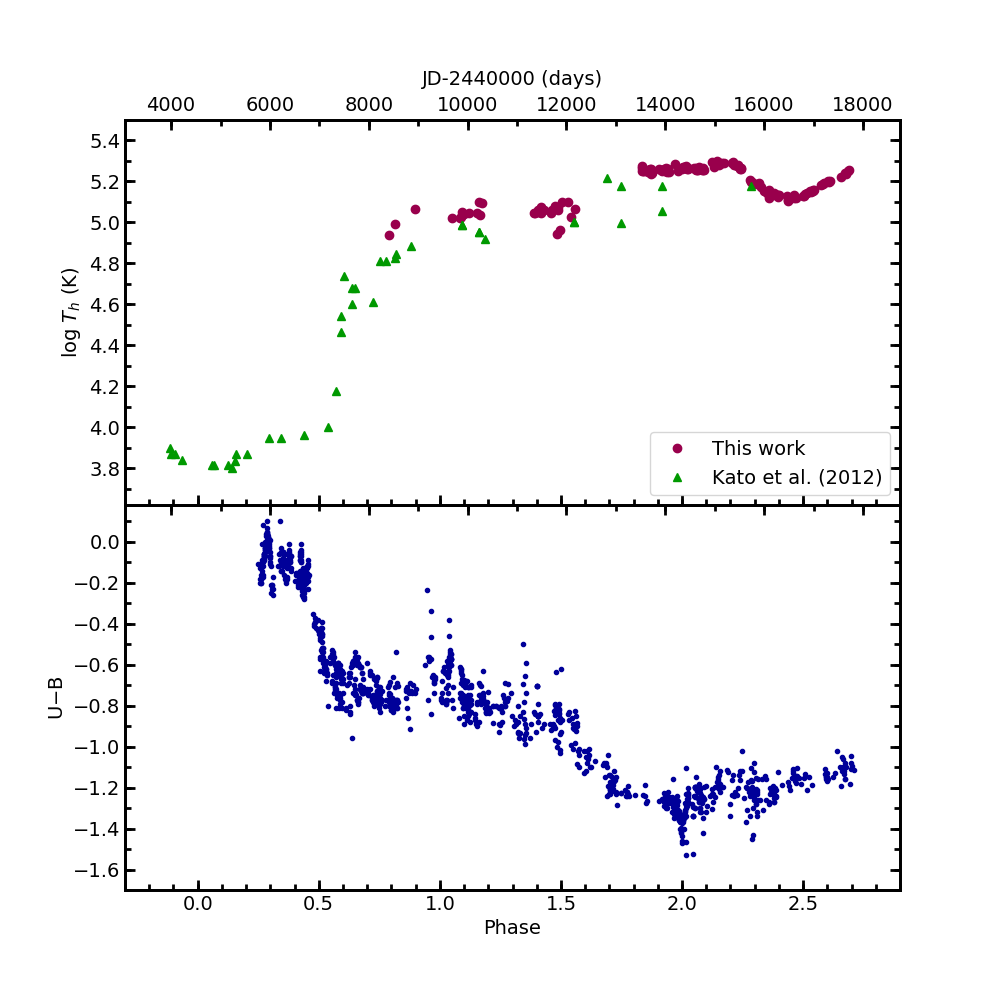}
\caption{Evolution of the temperature of the hot component (upper panel) and the evolution of the color U$-$B (lower panel).\label{fig:temp}}
\end{figure*}

The upper panel of Figure \ref{fig:temp} shows the derived temperatures (purple dots) as a function of time and phase for comparison with the temperatures from \citet[][green triangles]{2012ApJ...750....5K}. Although the observation epochs are different, the two sets of estimates are consistent. Nevertheless, our temperatures are slightly higher near phase 1. Considering the different methods used to derive temperatures, the agreement is reasonably good. Our results are also generally consistent with values from \citet{2011ARep...55..896T}, who reported an increase in temperature from $\sim$74,000~K in 1991 (JD 2448559) to $\sim$100,000~K in 1996 (JD 2450237). Our somewhat larger $T_h$ values -- $\sim$100,000~K for 1991 and $\sim$125,000~K for 1996 -- imply the same trend in the evolution of $T_h$ with time.

The large drop in $T_h$ at phase 2.4 is a result of the large illumination effect in H$\beta$ and He~I $\lambda$4471. During this epoch, the rise in the He~II $\lambda$4686 flux is smaller than the increases in He~I $\lambda$4471 or H$\beta$; thus the derived temperature declines. It is unlikely that the actual temperature of the hot component drops during this phase. Instead, $T_h$ probably remains close to 200,000~K.

The lower panel of Figure \ref{fig:temp} shows the evolution of the U--B color, consistent with an increase of the nebular continuum that dominates optical wavelengths from 1987. The increase in $T_h$ correlates well with the time evolution of the U--B color. The sharp rise in $T_h$ at phase 0.5 coincides with a dramatic drop in U--B, from U--B $\approx -0.2$ to U--B $\approx -0.7$. From phase 0.8 to phase 2.0, a gradual drop to U--B $\approx -1.2$ parallels the slow rise in $T_h$ from $10^5$~K to nearly $2 \times 10^5$~K. After phase 2, the U--B color gradually reddens to U--B $\approx -1.1$. During this period, the apparent $T_h$ changes due to the illumination effect. Although illumination is also visible in the broadband photometry, there is little evolution in the broadband colors during this period.

To check our $T_h$ measures, we consider the method of \citet{1994A&A...282..586M}:

\begin{equation}
\frac{T_h}{X_{max}}=1000~ ^\circ K/eV,
\end{equation}

\noindent where $X_{max}$ is the highest observed ionization potential. The highest ionization potentials in the observed spectra belong to [Fe~VII] $\lambda$6087 (ionization potential 99 eV) and O~VI $\lambda$6830 (ionization potential 114 eV), which give temperatures of 99,000~K and 114,000~K, respectively. These temperatures represent lower limits, due to the lack of higher ionization potential ions observable at optical wavelengths. Similarly, the lack of a detectable [Fe X] 6374 emission line (ionization potential 233 eV) place a firm upper limit of \mbox{$T_h \le$ 233,000 K}. On our spectra, the first appearance of [Fe~VII] and Raman-scattered O~VI occurs at JD 2453525 (phase $\sim$1.8, a little over a year after a gradual decline in U--B from $-$1.0 to $-$1.3), which coincides with an increase in $T_h$ from a little more than 100,000~K to close to 200,000~K from the H and He line fluxes \citep[see also][]{tat2018}. Thus the rise in $T_h$ during this epoch coincides with an increase in the ionization of the nebula.

\subsection{Luminosities of the hot component}
\label{luminosities}
Estimating the luminosity of the hot component from the emission lines also relies on simple assumptions. If the hot WD radiates as a blackbody and the ionized nebula absorbs {\it all} of the H-ionizing or He$^+$-ionizing photons, then predictions from case B recombination theory together with the observed H$\beta$ or He~II $\lambda$4686 fluxes yield the luminosity of the hot WD. If the nebula only absorbs a fraction of the ionizing photons, then the derived luminosity is a lower limit to the true luminosity. In this case, it is necessary to adopt a more detailed model to infer the luminosity of the hot WD \citep[e.g.,][]{1996ApJ...471..930P,1998ApJ...501..339P}. Even if the nebula is optically thick to ionizing photons, the simple blackbody + case B recombination model fails when the central star is a WR star \citep[e.g.,][]{1987ApJS...63..965H}. Comparisons with detailed models for WR stars then provide more accurate estimates of the luminosity \citep[e.g.,][]{2001AJ....122..349K}. To provide continuity with previous studies and to understand the trends in the evolution from our data, we follow the standard blackbody model and note where our assumptions might be inaccurate.

Following \citet{1983PhDT.........8K} and \citet{1997A&A...327..191M}, we estimate the luminosity of the hot component from the dereddened line fluxes of H$\beta$ and He~II $\lambda$4686. We assume that the hot component emits as a blackbody with $L_{h}=4\pi r_{h}^{2} \sigma T_{h}^{4}$, where $L_{h}$, $r_{h}$ and $T_{h}$ are the luminosity, radius and temperature of the hot component, respectively, and $\sigma$ is the Stefan-Boltzmann constant. The observed line fluxes are related to the number of ionizing photons ($f$), the emission coefficient ($K$) and the recombination coefficient ($\alpha$). For blackbody emission, $f$ is a function of $r_{h}$, $T_{h}$ and the energy required for ionizing H or He$^{+}$. Combining these, for H$\beta$, we derive:

\begin{equation}
L_{h}(H_{\beta})=\frac{4\pi d^{2}\alpha_{H}\sigma T_{h} F(H_{\beta})}{f_{H}p K_{\beta}}, 
\end{equation}

\noindent where $d$ is the distance to the source, $F(H_{\beta})$ is the observed dereddened flux in H$\beta$ and \mbox{$p=1.52\times10^{11}~ photons~ cm^{-2}~ s^{-1}~ K^{-3}$} is the photon emission constant. Finally, setting the values $\alpha_{H}=1.43\times10^{-13}~ cm^{3}s^{-1}$ and $K_{\beta}=0.66\times10^{-25}~ erg~ cm^{3}s^{-1}$, corresponding to a typical electron density $N_e \approx 10^6$~cm$^{-3}$ and electron temperature $T_e=20,000$ K, we obtain

\begin{equation}
L_{h}(H_{\beta})=\frac{2.5\times10^7 \left(\frac{d}{kpc}\right)^{2}T_{h} F(H_{\beta})}{f_{H}}, \label{lum1}
\end{equation}

\noindent where $L_{h}(H_{\beta})$ is in units of solar luminosity. Adopting different values for $n_e$ and $T_e$ yield 5\% to 10\% differences in the numerical coefficient. 

Analogously, for the luminosity from the He~II $\lambda4686$ flux, setting $\alpha_{He^{+}}=9.08\times10^{-13}~ cm^{3}s^{-1}$ and 
\mbox{$K_{He^{+}}=7.16\times10^{-25}~ erg~ cm^{3}s^{-1}$}, for a typical electron density $N_e \approx 10^6$~cm$^{-3}$ and electron temperature $T_e$ = $20,000~ K$, we obtain

\begin{equation}
L_{h}(He^{+})=\frac{1.5\times10^7 \left(\frac{d}{kpc}\right)^{2}T_{h} F(He^{+})}{f_{He^{+}}}. \label{lum2}
\end{equation}

For a blackbody, $f_{H}$ and $f_{He^{+}}$ are simple functions of the temperature of the hot WD derived in the previous sub-section and the energy required to ionize $H$ and $He^{+}$ (\mbox{$13.6$ eV} and $54.4$ eV, respectively). $f_{H}$ and $f_{He^{+}}$ vary from 0 for very low $T_{h}$ ($\sim$10$^{3}$ K) to 1 for very high $T_{h}$ ($\sim$10$^{5}$ K). For an adopted distance \citep[4.7~kpc;][]{2012ApJ...750....5K} and temperature and a measured dereddened flux from H$\beta$ or He II $\lambda$4686, the luminosity follows from these two relations.  

\begin{figure*}
\centering
\includegraphics[trim=0 20mm 0 6mm,width=\textwidth]{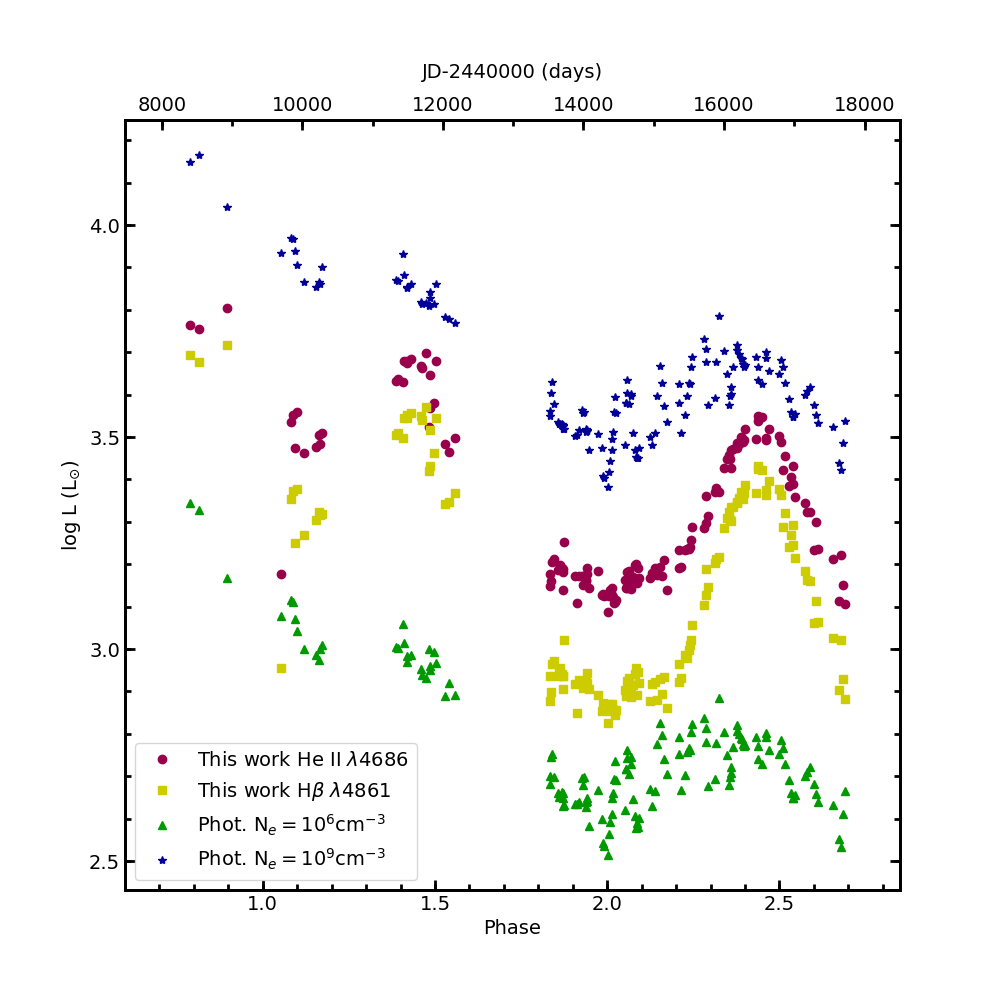}
\caption{Luminosities from He II $\lambda4686$ (purple dots) and H$\beta$ (yellow squares) line fluxes vs time and phase, in comparison with luminosities calculated with the method of \citet[][Phot.]{1994A&A...282..586M} for two different electronic densities. \label{fig:MNlum}}
\end{figure*}

Figure \ref{fig:MNlum} shows our results for the He II $\lambda4686$ (purple dots) and the H$\beta$ (yellow squares) line fluxes. The long-term evolution in $L_h$ is similar for the two measurements: at phases near 1.0, $L_h$ drops and then recovers to a slightly lower maximum level at phase 1.4. Following the maximum, $L_h$ falls to a distinct minimum, maintains a roughly constant value for nearly half the orbit, then rises to a clear peak at phase 2.4, and falls once again to a low luminosity. The luminosity before and after this latest peak is basically the same: for the He II $\lambda$4686 line the median luminosity before the peak is 1,461 L$_\odot$, while after it the last value is 1,275~L$_\odot$; for H$\beta$, the median luminosity is 813 L$_\odot$ before the peak and 761 L$_\odot$ after it. 

To test these results, we applied the method of \citet{1994A&A...282..586M} to the broadband optical photometry. In this approach, the nebula absorbs all of the high energy photons from the hot WD. \citet{1994A&A...282..586M} adopt a density structure for the ionized nebula and use a photoionization code to predict the optical continuum from the hot WD and the surrounding nebula. Integrating over the optical passbands yields a set of bolometric corrections for the system. The luminosity of the hot WD then follows from 

\begin{equation}
\frac{L}{L\odot}=\left(\frac{d}{10~ pc}\right)^{2} 10^{0.4(M^{\odot}_{bol}-m_{V}+BC_{V}+A_{V})}, 
\end{equation}

\noindent where $M^{\odot}_{bol}=4.64$ is the bolometric magnitude of the Sun, $BC_{V}$ is the bolometric correction in the V$-$band and $A_{V}=0.93$. Similar equations for U and B bands yield other estimates of the luminosity.

To employ the \citet{1994A&A...282..586M} method, we must (i) estimate the contribution of the RG to the optical flux and (ii) adopt an electron density for material near the hot WD. For the V magnitude of the RG, a reasonable estimate is $V$ = 13.15, the observed brightness when the RG occults the WD and surrounding nebula near phase 2. To check this estimate, we derive an alternate V-band brightness from published IR magnitudes in \citet{2011ARep...55..896T}, optical colors for M5$-$M6 III stars in \citet{1998PASP..110..863P}, and 1 magnitude of optical extinction. The resulting magnitude, $V$ $\approx$ 13.26, is consistent with our assumption and suggests the RG contributes about half of the optical flux at phases 1.5 and 2.5.

\citet{1994A&A...282..586M} consider two options for the electron density. In symbiotics with short orbital periods of 2--3 years or less, the electron density is high, N$_{e}=10^{9}$~cm$^{-3}$. Systems where the RG is a Mira variable have orbital periods of decades and lower densities, N$_{e}=10^{6}$~cm$^{-3}$. Although PU Vul has some characteristics of short-period systems (e.g., a lack of warm dust near the RG), its long orbital period is more characteristic of symbiotics with Mira variables. For completeness, we derive the luminosity in the high and low density limits.

Figure \ref{fig:MNlum} compares the luminosities derived from the emission lines and the optical photometry.  Luminosities from the \citet{1994A&A...282..586M} method bracket the $L_h$ inferred from the emission lines, with the high density limit for the electron density yielding the larger $L_h$ estimate.  The offset in $L_h$ between the nebular (H$\beta$ and He~II) and photometric techniques is clearly a function of the adopted electron density. Apparently, an intermediate density between the two adopted values (e.g., $N_e \approx 3 \times 10^7$~cm$^{-3}$) would yield a photometric $L_h$ more similar to the emission line result. The smaller variation of the photometric $L_h$ due to the illumination effect is due to the contribution of the RG. In PU Vul and other symbiotics where the hot WD has $T_h \approx 2 \times 10^5$~K and $L_h \gtrsim$ 1000~L$_\odot$, the WD contributes no flux to the optical continuum. When the RG also contributes little or no flux, the optical continuum flux (and hence the UBV brightness) increases in step with the line flux \citep{1984ASSL..112.....A,1989agna.book.....O}. In PU Vul, however, the RG emits roughly half of the V flux; thus the rise in the photometric $L_h$ at phase 2.4 is smaller than the rise in the $L_h$ derived from the emission lines.

Overall, our data suggest a clear drop in $L_h$ at orbital phases 0.8--2.8. At the start of our observations, $L_h \approx$~6,500~L$_{\odot}$. Following the eclipse at phase 1, $L_h$ falls to roughly 3,000~L$_{\odot}$. The subsequent rise at phase 1.4 is due to the illumination effect and therefore is not a true measure of the hot component luminosity. After the illumination peak, the luminosity falls to roughly 1,000~L$_{\odot}$. Except for a second rise due to illumination of the RG wind, the luminosity maintains this low level from phase 1.9 to phase 2.8. This drop in luminosity is in agreement with \citet{2011ARep...55..896T}, who observed a decline by a factor 10 in the time period 1992--2008 (phases $\sim$0.8--2.1).  

\begin{figure*}
\centering
\includegraphics[trim=0 20mm 0 6mm,width=\textwidth]{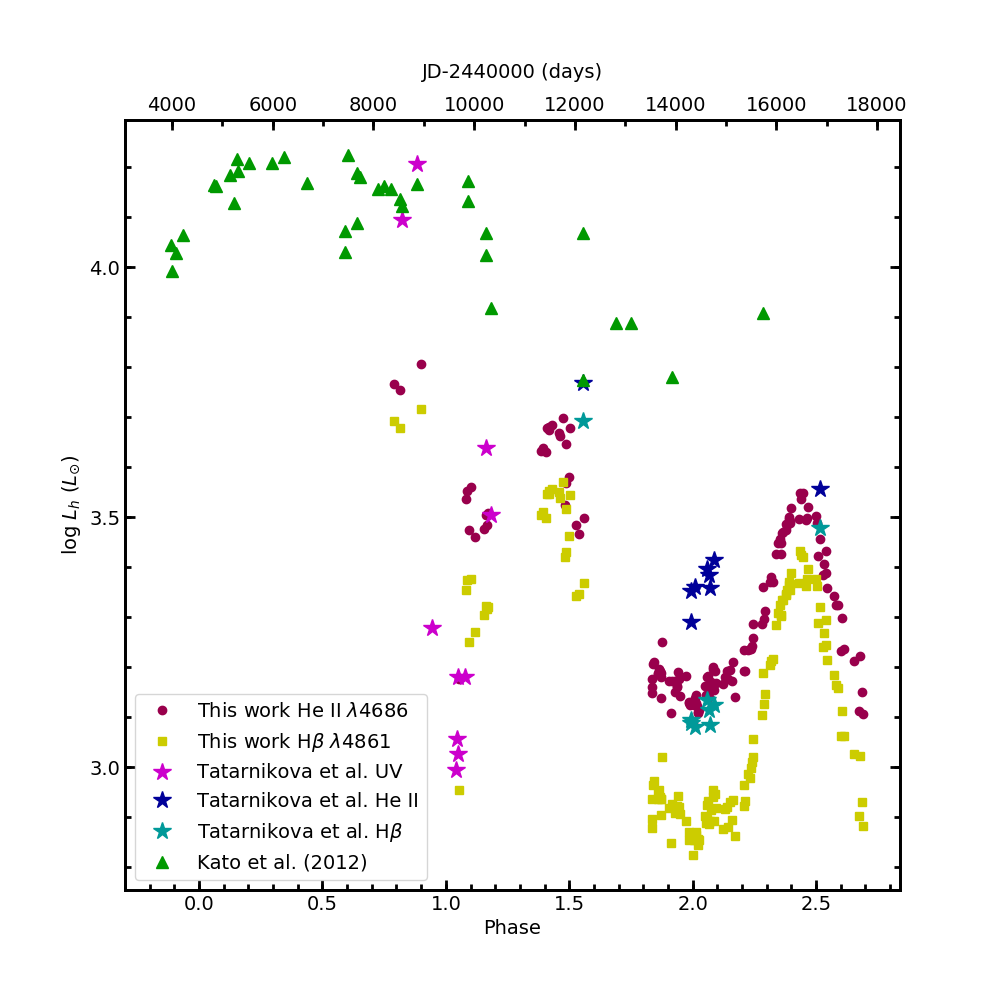}
\caption{Luminosities from He II $\lambda4686$ (purple dots) and H$\beta$ (yellow squares) line fluxes vs time and phase, in comparison with luminosities from \citet[][pink stars, UV data from JD 2448558 to 2450343]{2009ARep...53.1020T}, \citet[][light and dark blue stars, optical data from JD 2452167 to 2454763]{2011ARep...55..896T}, \citet[][light and dark blue stars, optical data from JD 2456857]{tat2018} and \citet[][green triangles]{2012ApJ...750....5K}.\label{fig:lum}}
\end{figure*}

In previous studies, \citet{2009ARep...53.1020T}, \citet{2011ARep...55..896T} and \citet{2012ApJ...750....5K} used a variety of techniques to infer $L_h$ for the hot WD in PU~Vul (Figure \ref{fig:lum}). In their analysis of archival IUE data from 1991 to 1996, \citet{2009ARep...53.1020T} adopted a smaller distance and a larger reddening than we consider here. To make a robust comparison with our analysis, we derived $L_h$ from their measured continuum fluxes at $\lambda$1300 and our adopted distance $d$ = 4.7~kpc and reddening \mbox{E(B--V)~$=$~0.3}. At 1991$-$1992 (JD $\sim$2448558$-$2448863), the derived $L_h$ is roughly a factor of two larger than our estimates, but they are comparable for the time period 1993$-$1996 (JD~$\sim$2449161--2450343). For the same set of IUE spectra, \citet{2012ApJ...750....5K} calculated similar luminosities than \citet{2009ARep...53.1020T} in 1991$-$1992 and more than two times larger in June 1996 (JD $\sim$2450237). Aside from IUE continuum data, \citet{2012ApJ...750....5K} estimated $L_h$ from (i) integrating the spectral energy distribution, (ii) the He~II $\lambda$1640 flux, (iii) the He~II $\lambda$4686 flux, and (iv) the photometric measure of \citet{1994A&A...282..586M}. \citet{2011ARep...55..896T} also derived $L_h$ from the He II $\lambda$4686 flux, but adopting again different parameters than the ones we considered. We estimated luminosities from their listed line intensities, using our distance and reddening values, and the method we described above. For the time period from 2001 (JD~2452167) to 2008 (JD~2454763), the resulting $L_h$ is almost a factor 2 higher than our estimates, for the two lines considered. We also employed our method to estimate the luminosity from fluxes and temperature for JD $\sim$2456857, in \citet{tat2018}. For this date and for both lines, H$\beta$ and He II $\lambda$4686, we found $L_h$ is higher again, but in a factor $\sim$1.3.  

Although various methods yield similar estimates for $T_h$ (Figure \ref{fig:temp}), there is fairly large disagreement in the derived luminosities. Early on in the outburst (before 1991, JD~$\sim$2448000), the various techniques employed by \citet{2012ApJ...750....5K} agree and suggest a roughly constant luminosity for the hot WD. As the system continued to evolve, the IUE data suggest either (i) a factor of two drop in $L_h$ \citep{2012ApJ...750....5K} or (ii) a factor of three drop in $L_h$ \citep{2009ARep...53.1020T}. Our measurements of the optical emission lines agree with the lower $L_h$ estimates from IUE data. Together with the optical analysis of \citet{2011ARep...55..896T}, our data and the optical photometry suggest another factor of four drop in $L_h$ from $\sim$2000 (JD $\sim$2451600) to the present day.

Several factors play a role in the different estimates for $L_h$. On IUE spectra, the continuum at $\lambda$1300 is weak and has an uncertain contribution from unresolved emission lines. The extinction correction at this wavelength is also much larger than the optical extinction. As noted earlier, the optical estimates (and UV measures based on emission lines) assume (i) the hot WD radiates as a blackbody and (ii)~the nebula absorbs all H-ionizing and He-ionizing photons. Given the uncertainties, it is impressive that the various techniques yield the same general trend of a hot WD with constant luminosity for some period of time followed by a gradual decline in $L_h$.

\section{Discussion}
\label{dis}
After the 1979 outburst, the hot component in PU Vul initially followed the evolution expected from a WD undergoing a thermonuclear runaway \citep{1984NASCP2349..305F,1984ApJ...279..252K,1986AJ.....91..563K,1996A&A...307..470N,2012ApJ...750....5K,2012BaltA..21..150S}. As noted by \citet{2012ApJ...750....5K}, the WD maintained a roughly constant luminosity of $10^4$ L$_\odot$ while the effective temperature increased from 7,000--8,000~K to more than $10^5$~K. During this period, the spectrum transformed from an A-F supergiant into a WR star and then a very hot star with strong emission lines from H, He, and other species. On our optical spectra, the highest ionization features ([Fe~VII] and Raman-scattered O~VI) appeared in 2005 (JD~2453525) and remained strong.  Although the evolution in the symbiotic nova AG Peg was much slower, it followed a similar pattern from the 1860s to the early 2000s \citep{1993AJ....106.1573K,2001AJ....122..349K}.

Starting in 1995--2000 (JD~$\sim$2449844--2451903), the derived luminosity of the hot WD began to drop (Figure \ref{fig:lum}). Despite a factor of 2$-$10 decline in $L_h$, the effective temperature of the hot WD remained roughly constant (Figure \ref{fig:temp}). This evolution runs counter to standard predictions, where the decline in $L_h$ occurs at roughly constant WD radius. However, the decline in $L_h$ is based on the assumption that the ionized nebula absorbs all of the H-ionizing and He-ionizing photons from the hot WD. Our observations of H and He line emission are not able to test this assumption.

The broadband light curves of PU Vul (bottom panel of Figure \ref{fig:flux_lc}) confirm some aspects of this picture. Aside from the eclipses, the initial decline of the UBVRI light curves is consistent with a hot WD evolving to higher $T_h$ at roughly constant luminosity. Later on, the light curves reveal two broad maxima at phases $\sim$1.4 and 2.4. These maxima correlate well with maxima in the emission line fluxes from H$\beta$, He~I $\lambda$4471, and He~II $\lambda$4686.  The similarity between the broadband light curves and the evolution of H and He emission line fluxes (3 upper panels in Figure \ref{fig:flux_lc}) suggests that most of the H and He emission is generated within the ionized, outflowing wind of the RG. The slight offset of the maximum from the expected phase 1.5 (and 2.5) is consistent with an elliptical orbit, e $\geq$ 0.16.  Together with optical pulsations \citep{2012BaltA..21..150S}, the relative amplitudes of the illumination effect in UBVRI and the emission lines suggest the RG is responsible for roughly half of the V-band light.

Before phase 2, the behavior of the He II $\lambda$4686 line (panel 2 of Figure \ref{fig:flux_lc}) implies formation in the expanding WR-type wind of the hot WD \citep[see also][]{1991MNRAS.252P..31T,1995IBVS.4251....1A,1995MNRAS.275..185K}. The appearance of broad He~II $\lambda$4686 on our optical spectra (before phase 1.15 and between phases 1.5 and 1.6--1.8) suggest the WR wind remained fairly strong throughout the first illumination peak at phase 1.4 (when narrow emission lines dominate the spectrum).  Somewhere between phases 1.6 and 1.8, the WR features fade away; the WD then gets hotter (Figure \ref{fig:temp}). The rise in the hot component temperature coincides with the appearance of the high ionization Raman-scattered O VI $\lambda$6830 line and a significant drop in the U$-$B color.  Several authors have noted a disappearance and re-appearance of He~II $\lambda$4686 during the second eclipse \citep[e.g.,][]{1996A&A...307..470N}, supporting the idea of formation close to the WD during this epoch. Although we do not have enough data around the second eclipse (phase 1) to verify the disappearance, the line is visible but weaker on 15 spectra acquired throughout the third eclipse (phase 2). This behavior is consistent with our contention that some He~II $\lambda$4686 emission comes from close to the hot WD, while the rest is within the ionized wind of the RG.

Aside from our picture, there are several alternative models for the time-variations in PU Vul.  \citet{2012ApJ...750....5K} considered a model consisting of a pulsating RG (average V~$\approx$~13.6), a partially ionized RG wind (V~$\approx$~14), a hot WD (which currently contributes little to the optical flux), and an ionized WD wind (V $\approx$ 12 during the second eclipse). Compared to our model, the RG and its wind are somewhat fainter, while the WD wind is somewhat brighter. Although this model provides a reasonable explanation for the behavior of PU Vul during the second eclipse, our analysis of the H and He lines demonstrates that emission from the ionized RG wind currently dominates emission from the WD wind. Our measurements of F(H$\alpha$)/F(H$\beta$) suggests this region has a modest optical depth, $\tau_\alpha \approx$ 5--10, in the H$\alpha$ line; the optical depth in the continuum is rather small. 

\citet{2011ARep...55..896T} proposed that an accretion disk around the hot WD contributes an observable amount of optical and UV flux. Although \citet{2012ApJ...750....5K} dismissed this hypothesis due to the lack of observational evidence, disk emission is a viable explanation for the UV and optical emission of some symbiotic stars \citep{1984ApJ...279..252K}. Within our set of spectra, there is little evidence for the broad emission lines expected from material orbiting the WD. Prior to phase 1.6--1.8, broad He~II emission has been interpreted as emission from a WR wind instead of a disk. 

Testing these proposals requires additional data analysis and better theoretical models. As we noted earlier, bolometric corrections for nebular models depend on the density of the RG wind near the hot WD \citep[Figure \ref{fig:MNlum};][]{1994A&A...282..586M}. In principle, the relative line fluxes of various intercombination and forbidden lines provide good constraints on the electron density \citep[e.g.,][and references therein]{1991AJ....101..637K,1993AJ....106.1573K}. In the optical, prominent [O~III], [Ne V], and [Fe VII] emission features can place good constraints on $N_e$ and $T_e$ \citep[e.g.,][]{1981A&A....99..177N,1982A&A...113...21N,1987A&AS...69..123N}. Coupled with O~III], C~III], and Si~III] lines on IUE spectra \citep[e.g.,][]{1984A&AS...56..293N,1986A&A...155..205N} and good reddening estimates, it should be possible to derive robust limits on the physical properties of the ionized nebulae throughout the outburst. For PU Vul, all of these lines are detected on UV and optical spectra; we defer an analysis of these features to a future study.

Placing these measurements in context requires a better understanding of the overall geometry of the RG wind.  \citet{1984ApJ...286..263T} analyzed different scenarios for an ionized nebula where a hot WD photoionizes the spherically symmetric wind of the RG. If the WD lies at a distance $a$ from the center of the RG, balancing recombination and ionization rates for a specified outward flow of neutral hydrogen atoms yields the shape of the surface of the resulting ionized nebula:

\begin{equation}
X=\frac{4\pi \mu^{2}m^{2}_{H}}{\alpha}a L_{ph}\left(\frac{\dot{M}}{\nu}\right)^{-2}, 
\label{xpar}
\end{equation}

\noindent where $\mu$ is the average mass of a particle in the wind, $m_{H}$ is the mass of an hydrogen atom, $\alpha$ is the recombination coefficient, $L_{ph}=f_{H}4\pi R_{h}^{2}pT_{h}^{3}=\frac{f_{H}p}{\sigma}\left(\frac{L_{h}}{T_{h}}\right)$ is the luminosity of the hydrogen ionizing photons emitted per second, $\dot{M}$ is the mass loss rate and $\nu$ is the terminal velocity of the wind. Depending on the $X$ parameter, the system may have three different geometries:
\begin{enumerate}
 \item $X<\frac{1}{3}$, for an ionization bounded nebula, surrounded by the neutral part of the wind.
 \item $\frac{1}{3}<X<\frac{\pi}{4}$, for a cone shaped ionized nebula, density limited in the outward direction, but still primarily ionization bounded.
 \item $X>\frac{\pi}{4}$, for a density bounded nebula, with a cone--like shadow zone as the neutral part of the wind and limited by the dense region near the cool star. 
\end{enumerate}
The emergent radio spectrum then depends on the parameter X. Originally developed for radio emission, the model is also valid for H~I Balmer lines and other emission lines. In section \ref{luminosities}, we described a method to derive luminosities assuming that the nebula absorbs all of the photons from the hot component. This approach represents the usual situation in symbiotic systems and corresponds to the first case of \citet{1984ApJ...286..263T}. For an optically thick nebula, our results suggest a rapid drop in $L_h$. Alternatively, the system could have developed a density bounded nebula (e.g., \mbox{$X > \frac{1}{3}$)}, where our approach underestimates the true $L_h$. If we set typical parameters for symbiotic stars \citep{1984ApJ...284..202S} in equation \ref{xpar} and typical parameters for PU Vul, we get: 

\begin{equation}
X=A~\frac{a}{7AU} f_{H} \frac{L_{h}}{1000L_{\odot}} \frac{150000K}{T_{h}} \left(\frac{10^{-7}M_{\odot}/yr}{\dot{M}}\frac{\nu}{10km/s}\right)^{2}.
\label{xpar_1}
\end{equation}

\noindent The chosen value for $a$ comes from using $P=13.4$ years and assuming a combined mass of 2 M$_{\odot}$ for the binary system \citep[e.g.][]{2003ASPC..303....9M} in Kepler's third law: $a=[P^{2} (M_{1}+M_{2})]^{1/3}=7.1$ AU. Then, setting $\mu=0.61$ for a nebula with H$^{+}$ and He$^{++}$ \citep[see, for example,][]{1958ses..book.....S}, $\alpha=1.43\times10^{-13}~ cm^{3}s^{-1}$ and $f_{H}=0.64-0.9$ from our calculations, we get $A\sim11.1-15.6$. According to equation \ref{xpar_1} the parameter X is basically A and corresponds to the third case in \citet{1984ApJ...286..263T}. Therefore, a better knowledge of the geometry of the nebula would allow a better determination for temperatures and luminosities of the hot component. 

Although published radio observations of PU Vul are not available, high spatial VLA observations covering a broad range of wavelength could establish the geometry of the nebula \citep[e.g.,][and references therein]{2003A&A...398..159G,2007A&A...471..825A,2015A&A...577L...4V}. If radio data are capable of setting the $X$ parameter in the \citet{1984ApJ...286..263T} model, then a combination of optical emission line and radio continuum fluxes yield better limits on $T_h$ and $L_h$.

In addition to the \citet{1984ApJ...286..263T} model, detailed models of the photoionized RG wind can also help to constrain $T_h$ and $L_h$ \citep[e.g.,][]{1996ApJ...471..930P,1998ApJ...501..339P}. In these calculations, optical and UV permitted, intercombination, and forbidden lines provide direct constraints on $T_h$ and $L_h$ as a function of the mass loss rate from the RG. With limits on $\dot{M}$ from radio data, it should be possible to learn whether the nebula has become density bounded or the hot WD has faded. Having a clearer picture of the evolution of the hot WD might teach us something new about thermonuclear runaways in symbiotic novae.

\section{Conclusions}
We have analyzed 32 years of optical spectroscopic observations for the symbiotic binary PU Vul. Together with the optical broadband light curve, the evolution of H~I, He~I, and He~II emission lines reveal a clear illumination effect, where the hot WD ionizes the expanding wind of the RG. The offset of the illumination peak from mid-eclipse points to an eccentric orbit with e $\geq$ 0.16. In the simplest interpretation of a nebula which absorbs all ionizing photons from the hot WD, the evolution of the H$\beta$ and He~II $\lambda$4686 emission lines imply an increase in effective temperature (from $T_h \approx$~80,000~K to $T_h \approx$ 200,000~K) and a corresponding decrease in luminosity (from $L_h \approx 5-10 \times 10^3$ L$_\odot$ to $L_h \approx 10^3$ L$_\odot$). Alternatively, the ionized nebula may have evolved from radiation bounded to density bounded. Radio observations covering a broad wavelength range, detailed analyses of optical and ultraviolet forbidden lines, and comprehensive photoionization calculations of the RG wind would test these conclusions.

\section*{Acknowledgements}
We acknowledge a generous allotment of telescope time on the FLWO 1.5-m telescope and the NOAO 0.9-m telescopes. We thank various remote observers on the FLWO 1.5-m. This paper uses data products produced by the OIR Telescope Data Center, supported by the Smithsonian Astrophysical Observatory. VC is supported by the Consejo Nacional de Investigaciones Cient\'ificas y T\'ecnicas (CONICET), Argentina. DC and SYS are supported by the Slovak Research and Development Agency under the contract No. APVV-15-0458 and by the Slovak Academy of Sciences grant VEGA No. 2/0008/17.




\bibliographystyle{mnras}
\bibliography{puvul} 






\bsp	
\label{lastpage}
\end{document}